\documentclass{PoS}
\pdfoutput=1
\usepackage{rotating,booktabs,amsmath,amssymb, multicol,url}
\usepackage[utf8]{inputenc}
\usepackage[numbers,sort&compress]{natbib}
\usepackage{natbibspacing}
\usepackage{ulem}

\title{Determination of the $N_f=12$ step scaling function using M\"obius domain wall fermions}

\ShortTitle{Determination of the $N_f=12$ step scaling function using M\"obius domain wall fermions}

\author{\speaker{Anna Hasenfratz}\\
        Department of Physics, University of Colorado Boulder, Boulder, CO, USA\\
        E-mail: \email{Anna.Hasenfratz@colorado.edu}
}
\author{Claudio Rebbi\\
        Department of Physics, Boston University, Boston, MA, USA\\
        E-mail: \email{rebbi@bu.edu}
}
\author{Oliver Witzel\\
        Department of Physics, University of Colorado Boulder, Boulder, CO, USA\\
        E-mail: \email{Oliver.Witzel@colorado.edu}
}
\abstract{We calculate the renormalized step scaling function for twelve fundamental flavors nonperturbatively by determining the gradient flow coupling on gauge field configurations generated with dynamical stout-smeared M\"obius domain wall fermions and Symanzik gauge action. Using Zeuthen, Symanzik, and Wilson flow we measure the energy density with three different operators. Our updated analysis is based on up to five volume pairs ranging from $L^4=8^4$ up to $32^4$. Predictions for the infinite volume extrapolated step scaling function based on different flows and operators are mutually consistent.  Our new results  confirm the previously observed significant discrepancy with  staggered fermion simulations in a wide range of the renormalized coupling.}

\FullConference{The 36th Annual International Symposium on Lattice Field Theory - LATTICE2018\\
		22-28 July, 2018\\
		Michigan State University, East Lansing, Michigan, USA.}

\begin{document}
\section{Introduction}
Non-abelian gauge theories are an important class of field theories with quantum chromodynamics (QCD) being one of the most prominent examples. Most properties of such gauge theories are determined by the gauge group (SU(3) for QCD) and the (light) fermion content, in QCD two fundamental Dirac flavors (up and down quarks). Varying the gauge group, the representation of the fermions, or the number of flavor changes the properties of such systems.

 Studying such systems as function of the fermionic degrees of freedom is of particular interest. Based on perturbative calculations, we expect that an SU(3) gauge system with fundamental flavors changes its nature from confining to conformal to infrared (IR) free as we increase the number of flavors. This change reflects how the gauge coupling $g$ depends on the energy scale $\mu$ and can be encoded in the $\beta$-function 
\begin{align}
\beta(g) = \frac{\partial g^2}{\partial \log(\mu)}
\end{align}
For an asymptotically free system, like QCD, the $\beta$-function is negative and exhibits only the trivial, Gaussian fixed point at $g=0$ where  the gauge coupling is relevant. At large $N_f$ where the system is IR free, the gauge coupling is irrelevant at the Gaussian fixed point and the $\beta$-function is entirely positive. In between is a range where the system exhibits conformal properties. Like a confining system, the $\beta$-function is negative at small $g$ but later develops  a second zero, corresponding to a conformal or IR fixed point. Near the lower onset of the conformal window this second fixed point occurs at strong gauge couplings, demanding non-perturbative calculations for reliable identification or confirmation. Establishing the lowest number of flavors exhibiting a conformal fixed point is important  because many composite Higgs models are based on near-conformal systems. Theories just {\it below}\/ the conformal window are expected to have near-conformal properties (e.g.~a ``walking'' gauge coupling) which are desired for building models which describe certain scenarios for physics beyond the Standard Model (SM).

Here we focus on the SU(3) gauge theory with twelve fundamental flavors and study its $\beta$-function numerically using lattice field theory simulations. Perturbative calculations at 2-, 3-, and 4-loop level predict a conformal fixed point, though the 5-loop $\overline{\text{MS}}$   result suggest that the IRFP, if it exists, is outside the range of convergence of perturbation theory.   There are several  nonperturbative lattice calculations that explore the infrared properties of this system \cite{Fodor:2011tu,Appelquist:2011dp,DeGrand:2011cu,Lombardo:2014pda,Fodor:2016zil,Hasenfratz:2016dou,Hasenfratz:2017mdh,Fodor:2017gtj,Hasenfratz:2017qyr}. To date, however, these results have led to different conclusions whether the SU(3) gauge system with $N_f=12$ fundamental flavors is indeed conformal, even predictions for the renormalized $\beta$-function are inconsistent over a wide range of the coupling $g$.\\

Trying to shed light on these discrepancies, we report updates on our numerical determination of the $N_f=12$ $\beta$-function \cite{Hasenfratz:2017mdh,Hasenfratz:2017qyr}. Our determination is based on the gradient flow (GF) step scaling function \cite{Luscher:2010iy,Fodor:2012td} that closely follows the negative of the continuum $\beta$-function
 \begin{align}
   \beta_s^{c}(g^2_c;L) = \frac{g^2_c(sL) - g^2_c(L)}{\text{log}(s^2)},
   \label{eq.beta_s}
\end{align}
with $s=2$. The finite volume normalized GF coupling 
\begin{align}
g^2_c (L) = \frac{128 \pi^2}{3(N_c^2 -1)}\;\frac{1}{C(c,L)} t^2 \langle E(t) \rangle
\label{eq.g2c}
\end{align}
is defined at flow time $\sqrt{8t} = cL$ where $c$ specifies the renormalization scheme and $C(c,L)$ is the tree-level normalization factor introduced in \cite{Fodor:2014cpa}. We calculate the step scaling functions on dynamical gauge field ensembles generated using stout-smeared M\"obius domain wall fermions (MDWF) and Symanzik gauge action \cite{Kaneko:2013jla,Morningstar:2003gk,Brower:2012vk,Luscher:1985zq,Luscher:1984xn}. Using \texttt{Grid} \cite{Boyle:2015tjk,GRID} we generate gauge field ensembles with $L^4$ volumes  and anti-periodic boundary conditions in all four dimensions at zero quark mass ($m_f=0$) for $L^4=8$, 10, 12, 14, 16, 20, 24, 28, and 32. Simulations are carried out using values of the bare coupling $\beta=6/g^2$ for $\beta=7.00$, 6.50, 6.00, 5.50, 5.20, 5.00, 4.80, 4.70, 4.60, 4.50, 4.40, 4.30, 4.25,  and 4.20 on all nine volumes.

Domain wall fermions (DWF) are formulated by adding a fifth dimension, $L_s$, and the 4d chiral modes are then projected onto the walls. DWF exhibit chiral symmetry which however gets slightly broken due to finite extent of the fifth dimension. The residual chiral symmetry breaking can be parametrized by an additive mass term, $m_\text{res}$. When increasing the strength of the coupling, it is well known that the residual chiral symmetry breaking grows.  We measure $m_\text{res}$ numerically and keep the residual chiral symmetry breaking under control and $m_\text{res}$ below $10^{-5}$ by increasing $L_s$ from 12 to up to 24 for simulations at strong couplings ($\beta \le 4.30$). The advantage of using expensive DWF compared to other formulations like, e.g.~Kogut-Susskind fermions \cite{Kogut:1974ag}, is that DWF preserve, even at finite gauge coupling, the full $SU(3) \times SU(3)$ flavor symmetry. Moreover, the effective gauge term generated by the fermions and the smearing is very small, mostly absorbed by the Pauli-Villars regulator of DWF. This leads to reduced cut-off effects and increases the region where  perturbative improvements are applicable. It is interesting to note that the expectation value of the plaquette normalized to 1, a good measure of the UV fluctuations of the gauge fields, is above 0.6  even at the strongest coupling in our DWF simulations.  We complement the good properties of domain wall fermions by using a fully $O(a^2)$ improved set-up:  Symanzik gauge action, Zeuthen flow \cite{Ramos:2015baa}, and  Symanzik improved  operator. Remaining discretization artifacts are further reduced by using the tree-level normalization in the definition of the gauge coupling (\ref{eq.g2c}) \cite{Fodor:2014cpa}.

\begin{figure}[tb]
  \centering
  \includegraphics[width=0.49\columnwidth,clip]{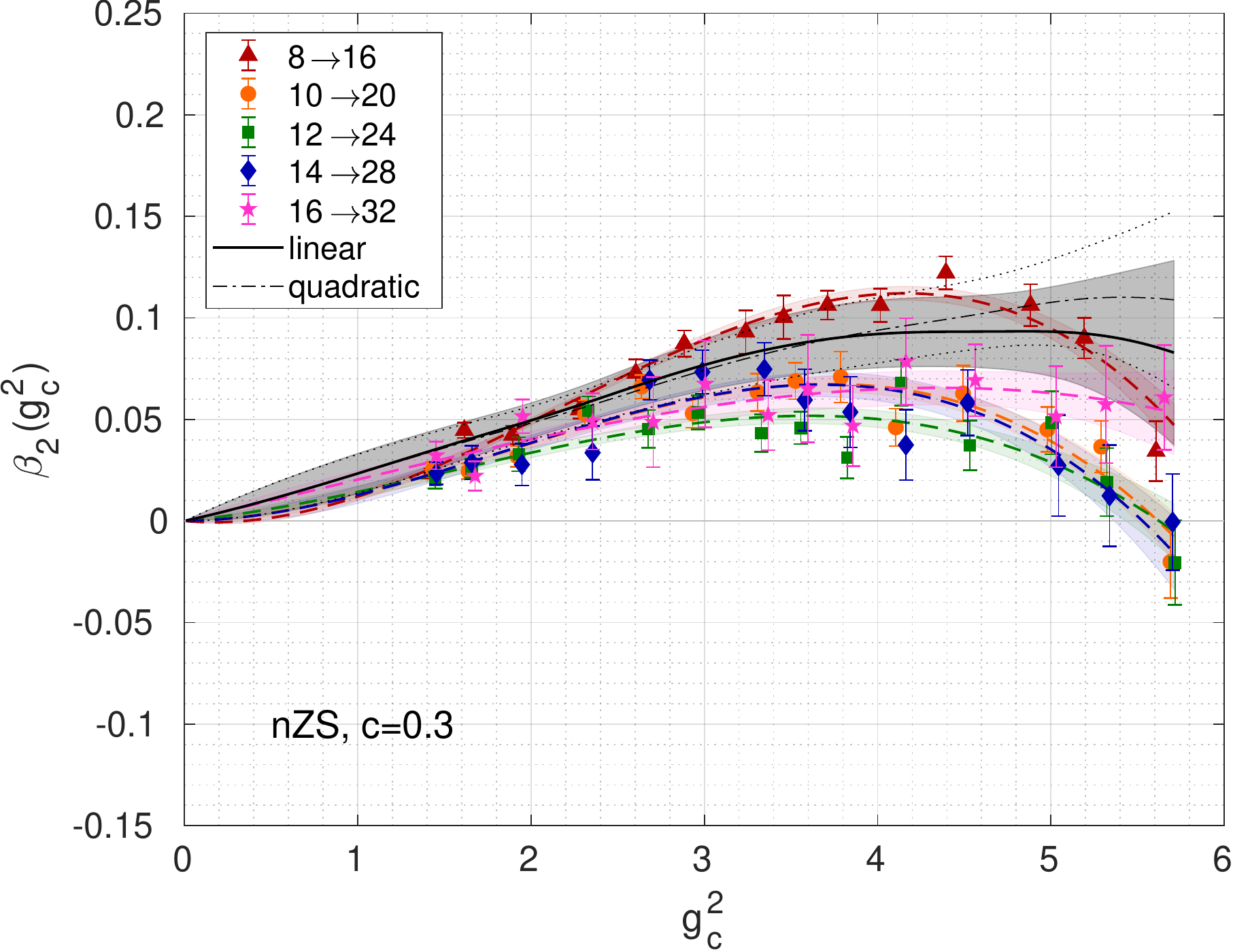}
  \includegraphics[width=0.49\columnwidth,clip]{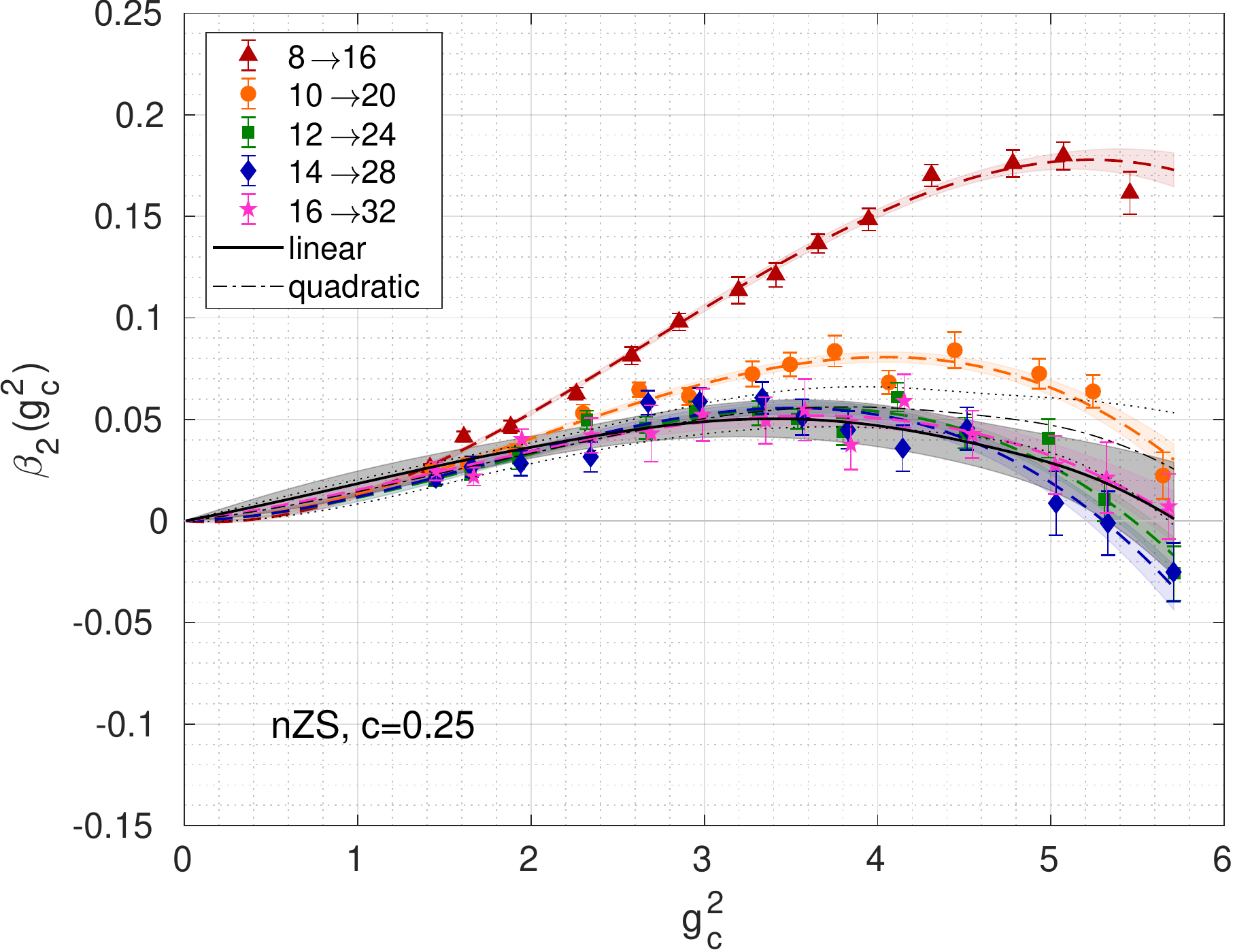}
  \caption{Non-perturbatively determined gradient flow step-scaling function with infinite volume limit extrapolation for the renormalization schemes $c=0.3$ (left) and $c=0.25$ (right). The colored data points are obtained using Eq.~(\protect\ref{eq.beta_s}) for the tree-level normalized (n) couplings determined using Zeuthen flow (Z) and the Symanzik operator (S). Colored bands correspond to the polynomial interpolation for each volume pair, while the solid black line with gray error band shows the linear infinite volume extrapolation and the black dashed-dotted line with $1\sigma$ uncertainty indicated by dotted lines shows the quadratic extrapolation.} 
  \label{fig.beta_s}
\end{figure}

Our updated results are shown in Fig.~\ref{fig.beta_s} using tree-level normalized (n) couplings obtained from Zeuthen flow (Z) using the Symanzik operator (S) for renormalization schemes $c=0.3$ (left) and $c=0.25$ (right). The continuum limit $L\to\infty$ step scaling function has a rather strong dependence on the parameter $c$. In the rest of this paper we will concentrate on $c=0.3$. Compared to our original analysis \cite{Hasenfratz:2017mdh,Hasenfratz:2017qyr}, we have added a fifth volume pair ($14 \to 28$) and increased statistics on many ensembles to have {\it at least}\/ $3500$ thermalized MDTU per ensemble. Further, we have revised our analysis strategy. For each volume pair, we obtain the tree-level normalized step scaling function (Eq.~(\ref{eq.beta_s})) shown by the colored data points in Fig.~\ref{fig.beta_s}. Next we interpolate these values in $g_c^2$ using a third order polynomial and draw the result using dashed lines with shaded error band in the same color as the data points. Because discretization errors of our data are extremely small and practically disappear for weak couplings, we constraint the intercept of the interpolation to vanish at $g_c^2=0$.\footnote{Theoretically this is only expected in the continuum limit; practically we cannot resolve a non-zero intercept for any of our volume pairs. Removing this constraint does not change any of the results presented here.}  In a second step, these interpolated step scaling functions are extrapolated to the continuum $L\to \infty$ limit using either a quadratic ansatz in $1/L^2$ for all five volume pairs (black dash-dotted line with dotted $1\sigma$ uncertainties) or a linear ansatz extrapolating the three largest volume pairs shown by the solid black line with gray error band. 
       Since we are calculating a renormalized quantity and extrapolate the GF step scaling function to the continuum $L\to \infty$ limit, we obtain a result which is expected to agree when compared to other determinations using the same renormalization scheme and is independent e.g.~of the fermion discretization. 
\\

In the next section we will present further details of our analysis, scrutinize it, and demonstrate the robustness of our findings. In Section \ref{sec.summary} we will address implications of our results, comment on the question of fermion universality before giving an outlook on ongoing and future work.

\section{Details of our gradient flow step scaling analysis}
As stated above, our preferred analysis is based on tree-level normalized couplings determined using Zeuthen flow and the Symanzik operator. Choosing the renormalization scheme $c=0.3$, we investigate other choices of our analysis.

\begin{figure}[tb]
  \centering
  \includegraphics[width=0.49\columnwidth,clip]{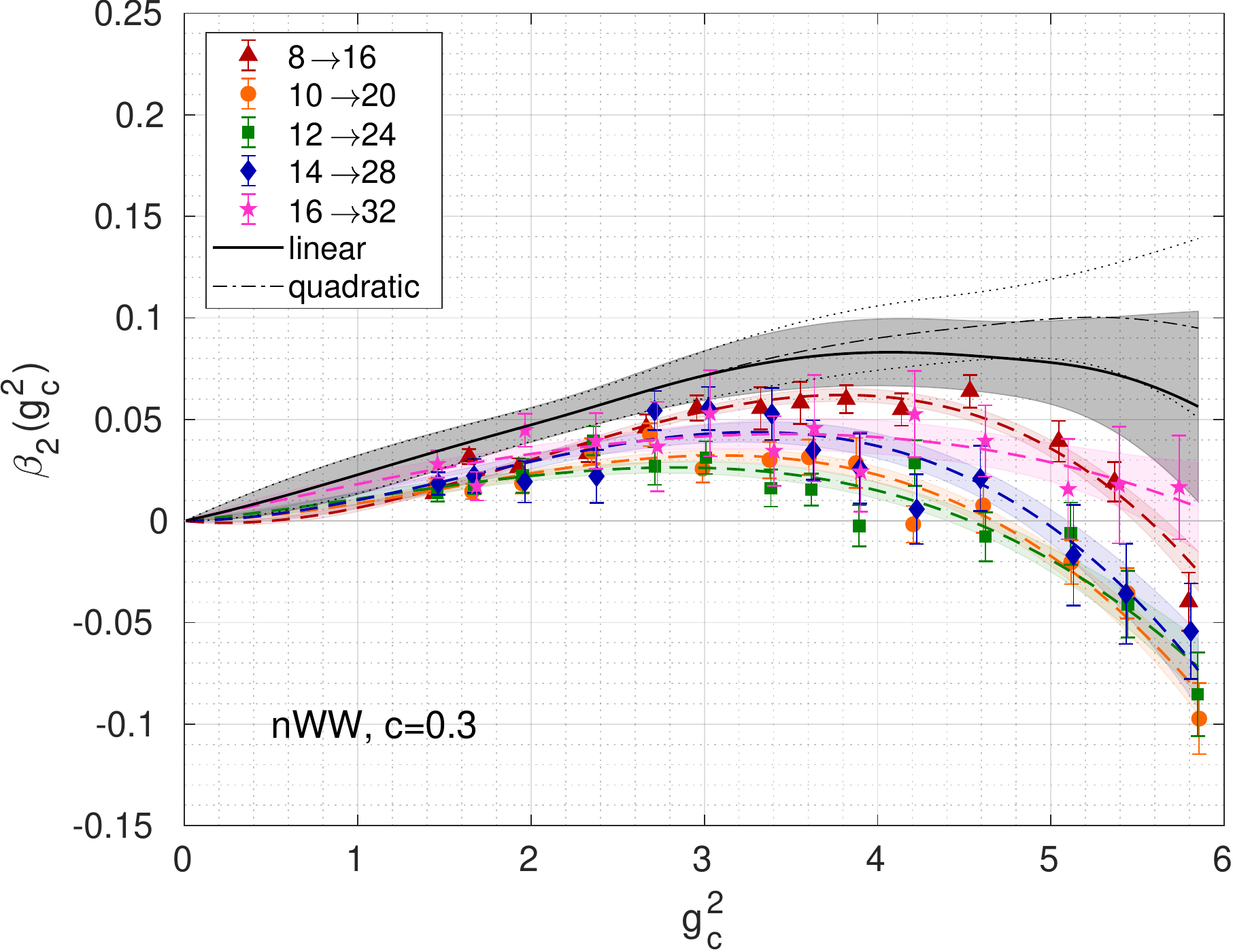}
  \includegraphics[width=0.49\columnwidth,clip]{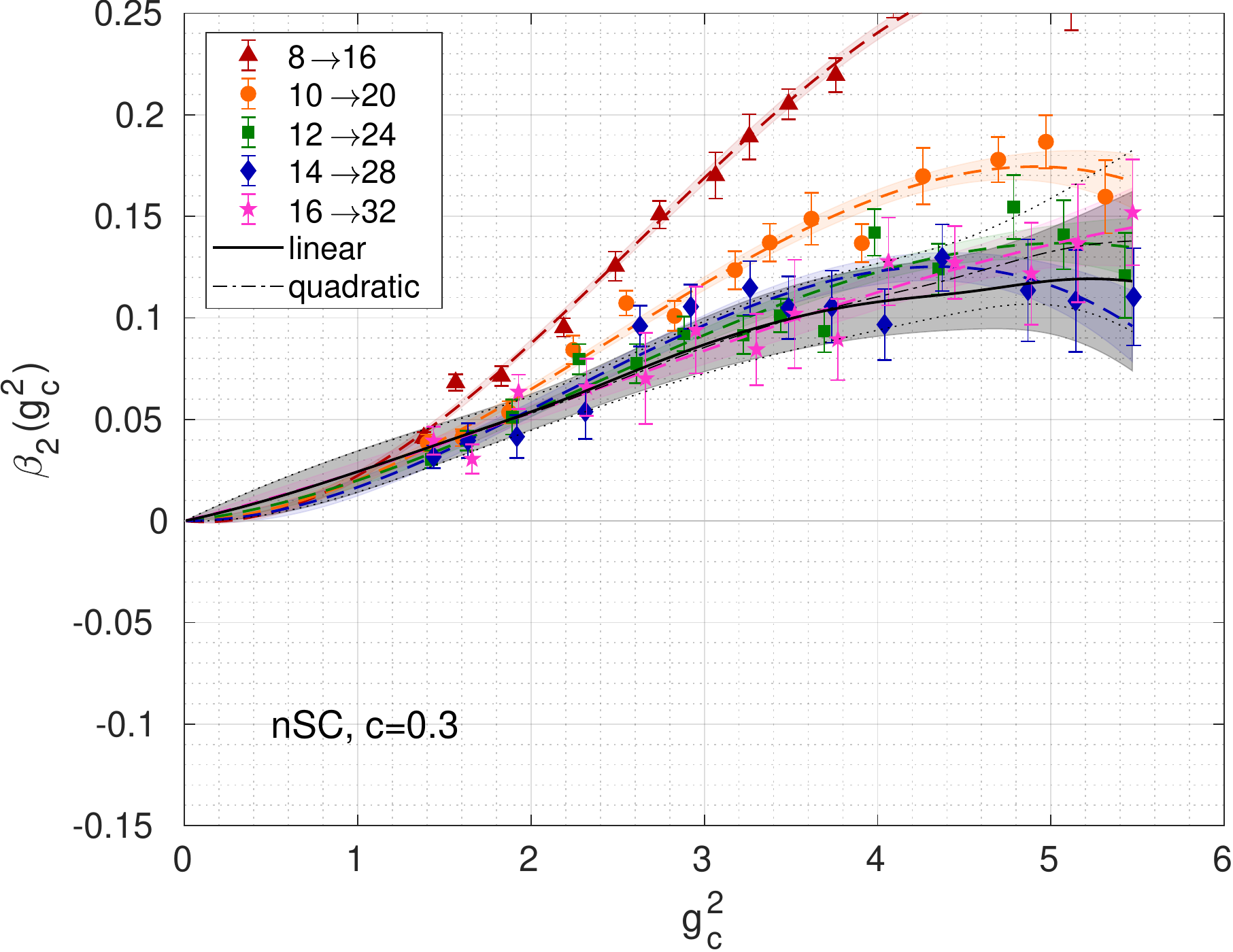}  
  \caption{Alternative determinations of the GF step scaling function applying tree-level normalization (n) to couplings determined using Wilson flow (W) with the Wilson operator (W) on the left or Symanzik flow (S) with the clover operator (C) on the right.}
  \label{fig.alt_flow_op}
\end{figure}

\begin{figure}[tb]
  \centering
  \includegraphics[width=0.49\columnwidth,clip]{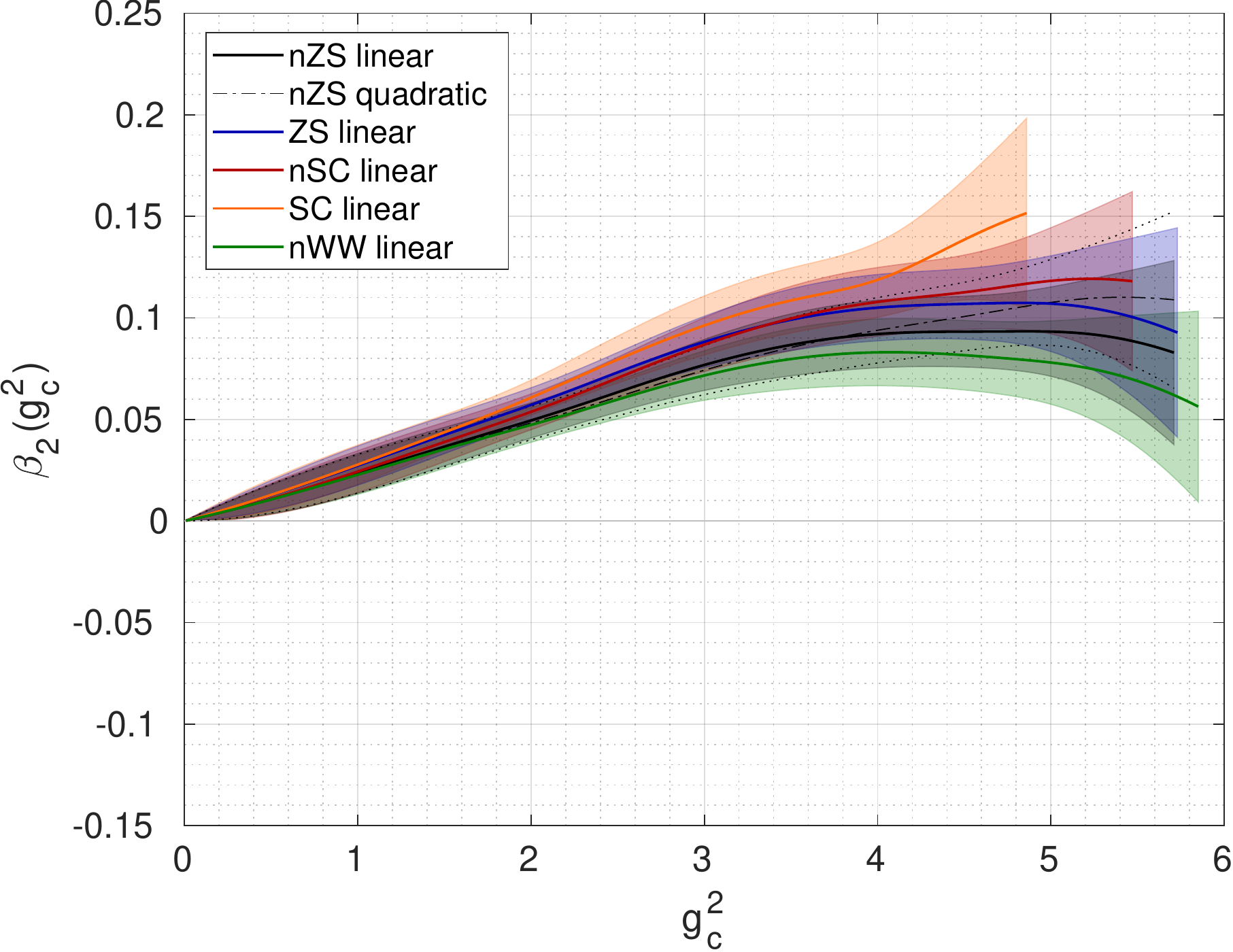}
  \includegraphics[width=0.49\columnwidth,clip]{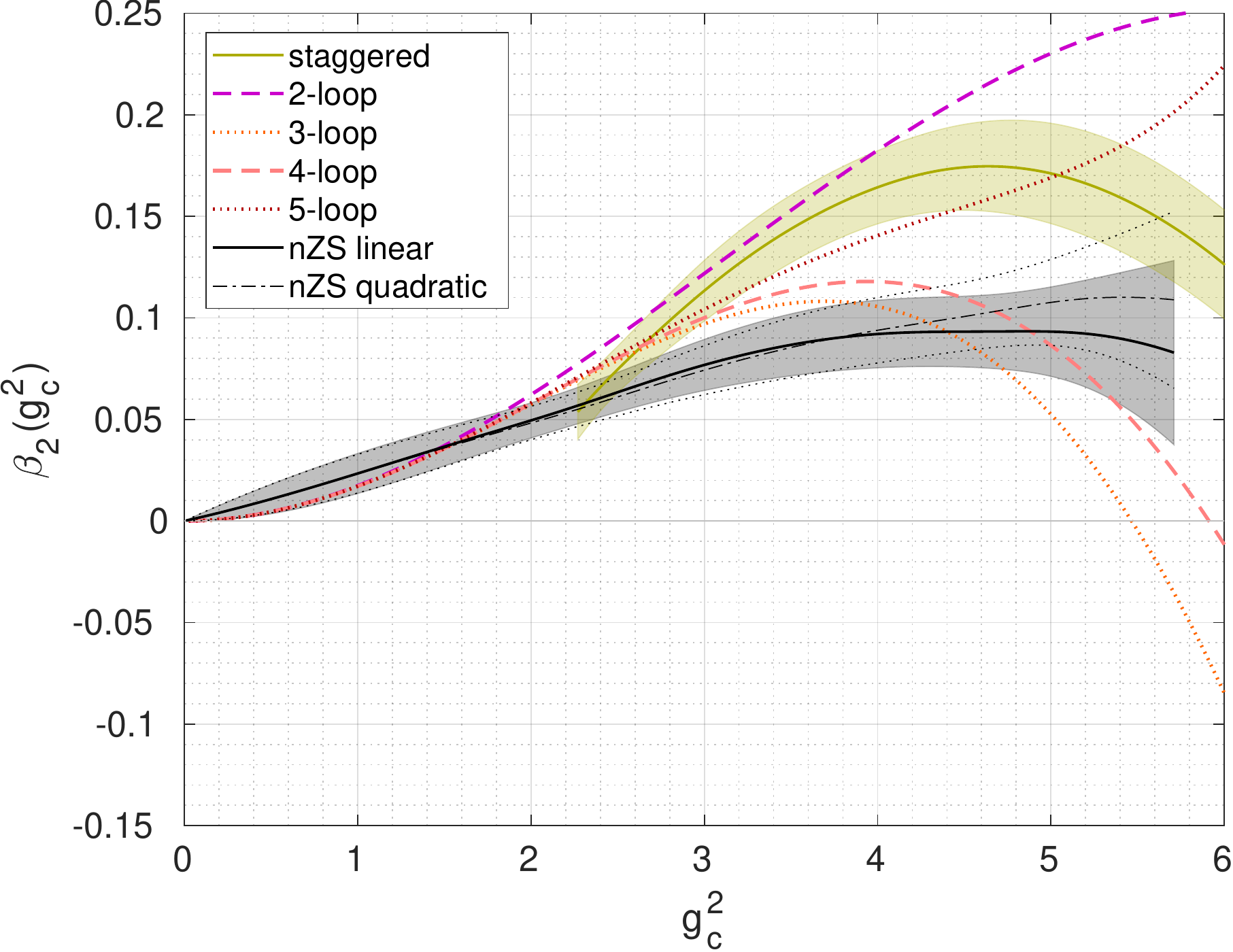}
  \caption{Left: comparison of continuum extrapolated GF step scaling functions obtained with or without tree-level normalization and different flows or operators. All shown curves overlap indicating agreement within $1\sigma$ statistical uncertainties. Right: comparison of our nZS result to perturbative prediction and the result based on a calculation with staggered fermions \cite{Hasenfratz:2016dou}.}
  \label{fig.compare}
\end{figure}

First we present in Fig.~\ref{fig.alt_flow_op} alternative determinations of the GF step scaling function using tree-level normalized Wilson flow (W) with the Wilson operator (W) (left) and  Symanzik flow with the clover operator (C) (right). Comparison with the left plot in Fig.~\ref{fig.beta_s} shows that while the data points for individual volume pairs change quite substantially, in particular for SC, the continuum $L \to \infty$ extrapolated result shows only minimal fluctuations, as can be seen in the comparison plot (Fig.~\ref{fig.compare}, left panel).

\begin{figure}[tb]
  \centering
  \includegraphics[width=0.49\columnwidth,clip]{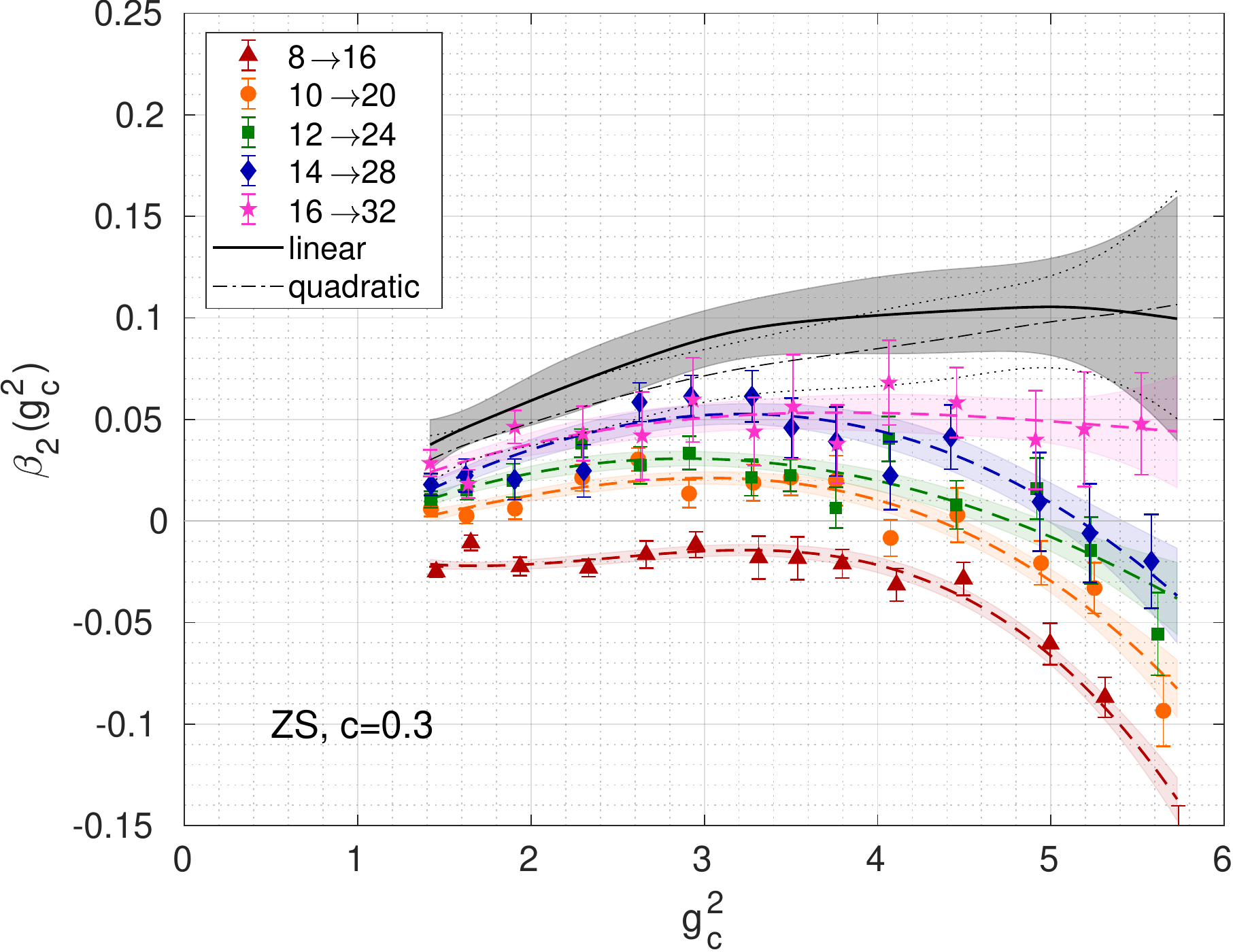}
  \includegraphics[width=0.49\columnwidth,clip]{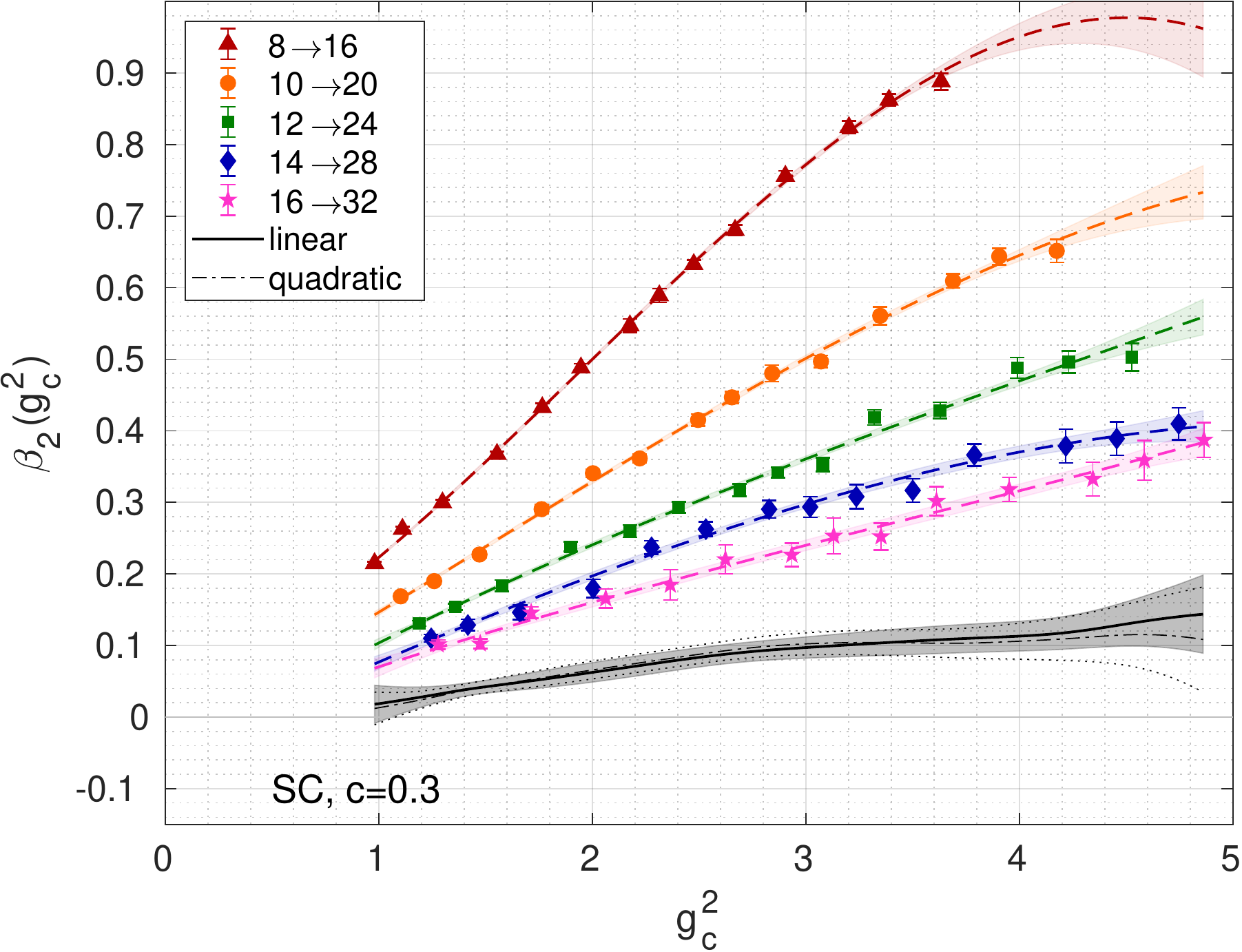}  
  \caption{GF step scaling function analyzed for ZS (left) and SC (right) without using the perturbative tree-level normalization. In both cases we removed the constraint on the intercept of the polynomial interpolation. Discretization effects for SC are large, forcing twice the range of the $y$-axis.}
  \label{fig.no_tln}
\end{figure}

Secondly, we demonstrate the effect of the tree-level normalization by showing in Fig.~\ref{fig.no_tln} the analysis without using the perturbative improvement. On the left, data obtained from Zeuthen flow and Symanzik operator are analyzed; on the right, Symanzik flow and clover operator are used. As can be seen by comparison to the left plot in Fig.~\ref{fig.beta_s} and the right plot in Fig.~\ref{fig.alt_flow_op}, the difference between the different volume pairs grows. This is a sign that tree-level normalization effectively reduces discretization effects and helps to improve the quality of the infinite volume extrapolation. We do not observe the breakdown of tree-level normalization reported in \cite{Fodor:2015baa} and attribute this  to the substantially smaller cut-off  effects observed with DWF. Despite discretization effects becoming significant without tree-level normalization, the continuum extrapolated results are  still in agreement within $1\sigma$ statistical uncertainty with our preferred analysis as shown in the left panel of Fig.~\ref{fig.compare}.

\begin{figure}[tb]
  \centering
  \includegraphics[width=0.49\columnwidth,clip]{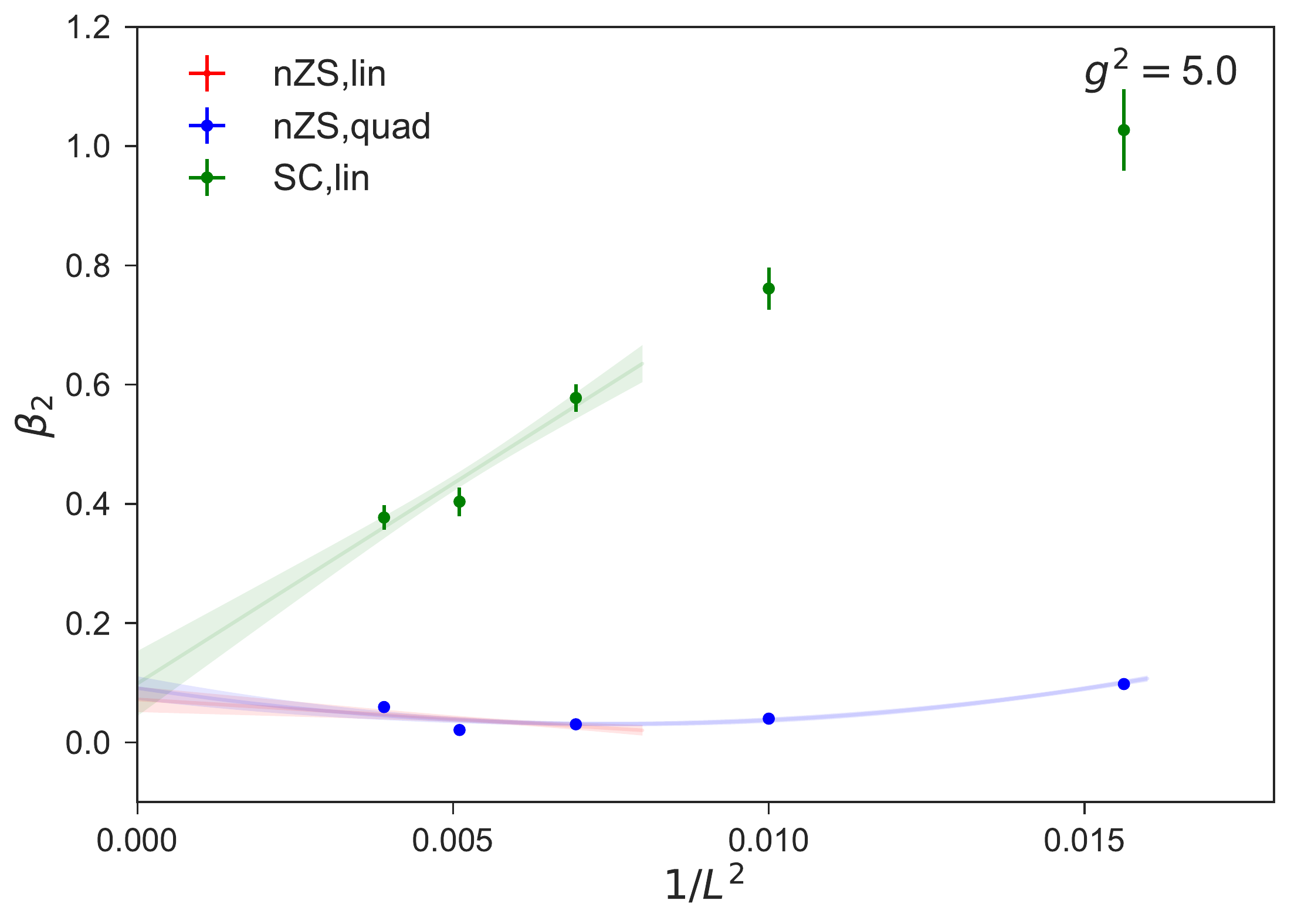}
  \includegraphics[width=0.49\columnwidth,clip]{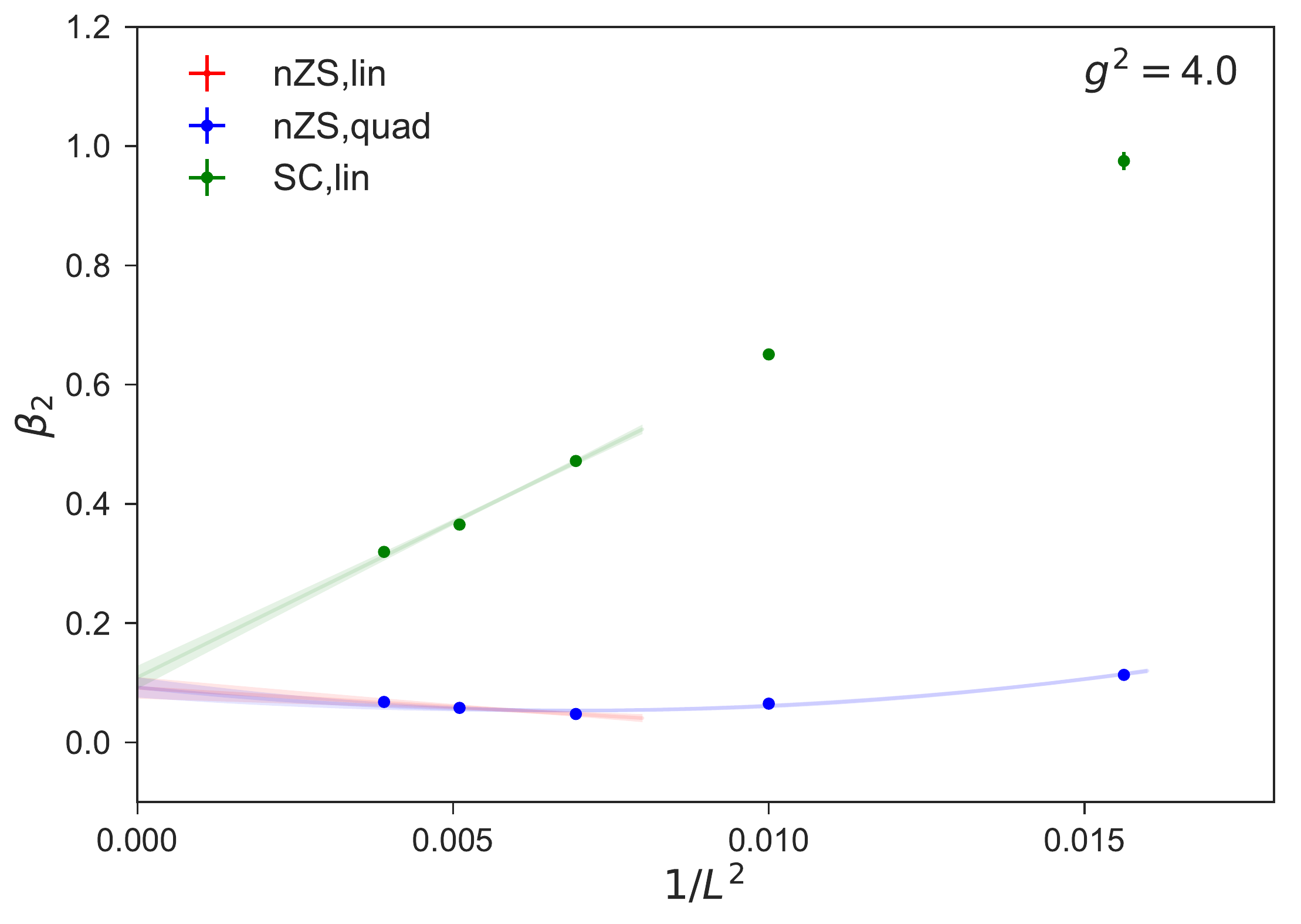}
  \\
  \includegraphics[width=0.49\columnwidth,clip]{images/Cont-extr_nZS_c03_g2_40}
  \includegraphics[width=0.49\columnwidth,clip]{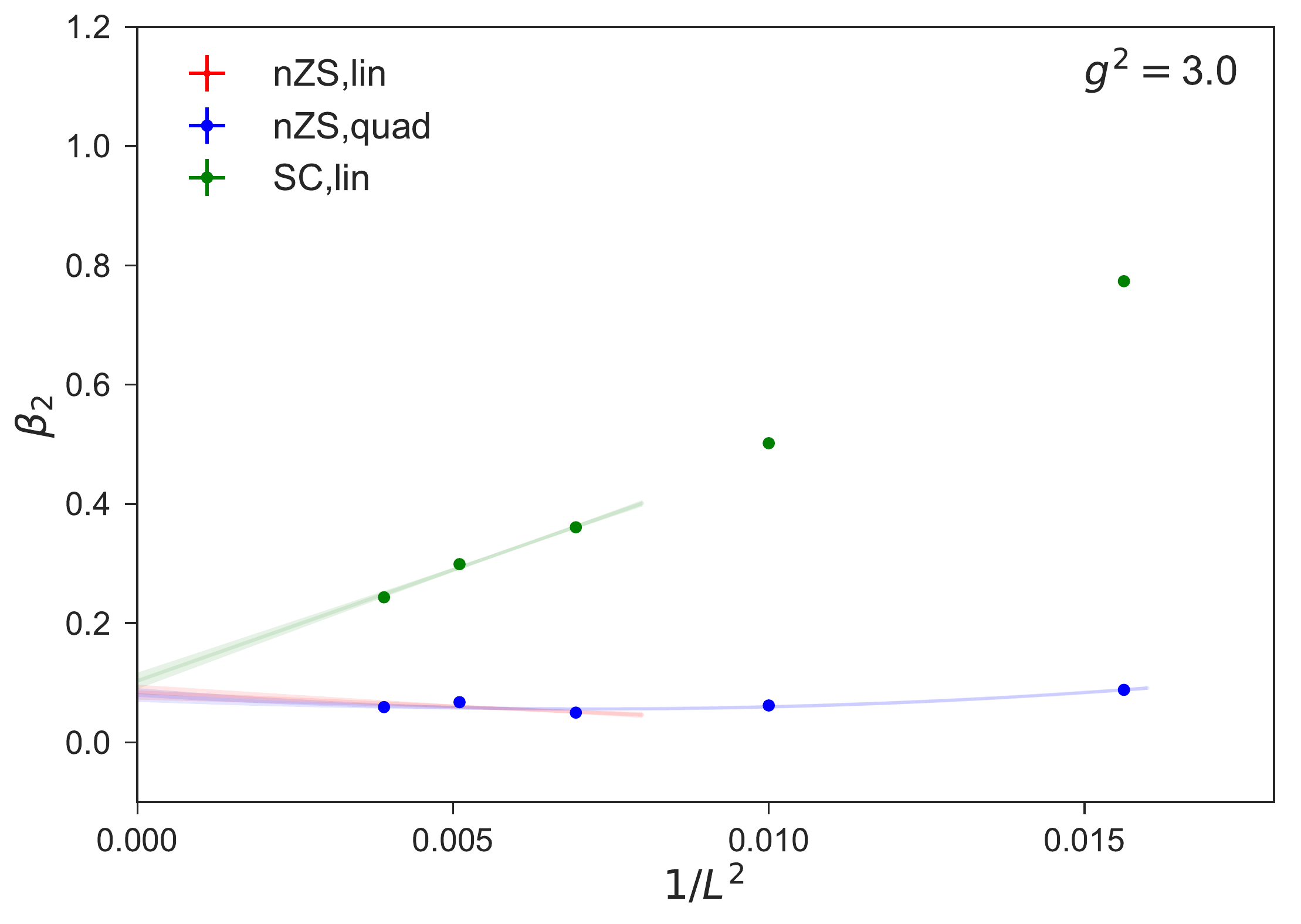}  
  \caption{Continuum $L\to \infty$ extrapolations at selected values of the coupling $g_c^2$. The smallness of discretization effects in our preferred nZS analysis is seen by comparison to SC data. Although the latter values spread over a much larger range, the fact that both determinations agree in the continuum is a sign of the quality of our data/set-up.}
  \label{fig.L_to_infty}
\end{figure}

Finally, we present details on the $L\to \infty$ extrapolations. In Fig.~\ref{fig.L_to_infty} we show for selected values of the squared renormalized gauge coupling, $g_c^2$, the extrapolations carried out by either using a linear fit ansatz to extrapolate the data points on the three largest volume pairs or an ansatz including an additional quadratic term in $1/L^2$ to fit data points from all five volume pairs. To demonstrate the smallness of the discretization effects in our preferred analysis based on tree-level normalized coupling obtained from Zeuthen flow with Symanzik operator, we also show the extrapolation without improvement using Symanzik flow and clover operator. Although the range of values in the extrapolation is quite different, the extrapolated results agree giving further credit to our data.

\section{Summary}
\label{sec.summary}

In Fig.~\ref{fig.beta_s} we presented our updated analysis of the GF step scaling using up to five different volume pairs and volumes up to $L^4=32$ to improve the continuum $L\to \infty$ limit extrapolation. Our analysis uses MDWF preserving the full $SU(3) \times SU(3)$ symmetry and a full $O(a^2)$ improved lattice set-up. Extrapolating the renormalized coupling to the continuum limit, we obtain a result which should be independent of lattice discretizations and is therefore expected to agree with other determinations (using the same renormalization scheme $c$) over the entire range of $g_c^2$ and not only at the IR fixed point.

Comparing  our result to the one presented in Ref.~\cite{Hasenfratz:2016dou} in the right panel of Fig.~\ref{fig.compare} or noting that Ref.~\cite{Fodor:2017gtj} finds a basically constant  $\beta$-function around 0.13 for $c=0.25$ and $g_c^2>6$,  we observe a stark discrepancy between our result and calculations using staggered fermions.   Further investigations are required to identify the source of the disagreement whether e.g.~universality of fermion formulations is violated in in conformal systems or whether systematic effects significantly alter the outcome. Discretization uncertainties might push the system out of the basin of attraction of the FP, resulting in different fermion formulations leading to different results.

In either case implications might be significant for  numerical investigations of conformal or near-conformal systems in the realm of beyond the SM physics. Due to the lower numerical costs,  staggered fermions have been the preferred choice for beyond the SM studies (see e.g.~\cite{Fodor:2016wal,Aoki:2016wnc,Appelquist:2018yqe,Brower:2015owo,Hasenfratz:2016gut}). To investigate possible scaling violation effects, we started  a spectral study of a near-conformal systems using MDWF \cite{Witzel:2018gxm}.  In parallel we continue our step-scaling calculations for SU(3) systems with $N_f=12$ or 10 fundamental flavors with the aim to push to even stronger couplings and/or improve the continuum extrapolation. Preliminary data for $N_f=12$ indicate an IRFP around $g_c^2=6$ for $c=0.25$ as is already suggestive in the left plot of Fig.~\ref{fig.beta_s}.

\section*{Acknowledgments}
We are very grateful to Peter Boyle, Guido Cossu, Anontin Portelli, and Azusa Yamaguchi who develop the \texttt{Grid} software library providing the basis of this work and who assisted us in installing and running \texttt{Grid} on different architectures and computing centers. Computations for this work were carried out in part on facilities of the USQCD Collaboration, which are funded by the Office of Science of the U.S.~Department of Energy, the RMACC Summit supercomputer, which is supported by the National Science Foundation (awards ACI-1532235 and ACI-1532236), the University of Colorado Boulder, and Colorado State University, and on \texttt{stampede2} at TACC  allocated under the NSF Xsede program to the project TG-PHY180005.
We thank  Fermilab,  Jefferson Lab, the University of Colorado Boulder, TACC, the NSF, and the U.S.~DOE for providing the facilities essential for the completion of this work.  A.H.~and O.W.~acknowledge support by DOE grant DE-SC0010005 and C.R. by DOE grant DE-SC0015845.

{\small
\bibliography{../General/BSM}
\bibliographystyle{apsrev4-1}
}



\end{document}